\documentclass[twoside]{article}
\usepackage{fleqn,espcrc2}

\usepackage{graphicx}

\newcommand{\AmS}{{\protect\the\textfont2
  A\kern-.1667em\lower.5ex\hbox{M}\kern-.125emS}}

\title{Intrinsic c-Axis Transport in 2212-BSCCO}

\author{P.J. Thomas, J.C. Fenton, G. Yang and C.E. Gough\address{Superconductivity Research Group, University of Birmingham, Edgbaston, Birmingham B15 2TT, United Kingdom}
        \thanks{Corresponding author email C.Gough@bham.ac.uk}}

\begin{document}
\begin{abstract}
We describe two experimental approaches to circumvent the problem of self-heating in IV measurements on small mesa samples of 2212-BSCCO.  Simultaneous dc and temperature measurements have been performed, allowing corrections for heating to be made. Short pulse measurements have also been made, where the IV characteristics and the mesa temperature can be measured on a $\mu $s time-scale enabling intrinsic IV characteristics to be derived, even in the presence of appreciable self-heating. Self-heating leads to an appreciable depression of the apparent energy gap and also accounts, in major part, for the s-shaped characteristics often reported at high currents. By correcting for the temperature rise, we derive the intrinsic temperature dependence of the tunnelling characteristics for crystals with a range of doping. Results are compared with recent theoretical models for c-axis transport in d-wave superconductors.
\vspace{1pc}
\end{abstract}

\maketitle

\section{Introduction}

It is now well established that highly anisotropic cuprates act as a series array of intrinsic Josephson junctions shunted by a quasiparticle conductivity and a significant capacitance\cite{schlenga}. Information can be obtained about the mechanism of c-axis transport by studying the temperature and voltage dependence of the tunnel current in the intrinsic junctions. In particular, the characteristics can be used to deduce the temperature dependence of the superconducting gap $\Delta$ and to provide information on the pairing symmetry. 

However, the interpretation is complicated by apparent gap suppression within samples of intrinsic junctions. This has been attributed to non-equilibrium effects \cite{suzuki1} or simple Joule heating. One method of minimising heating is to use stacks of intrinsic junctions (mesas) with small dimensions in order to reduce $I_\mathrm{c}$ and the resulting dissipation $\sim 2I_\mathrm{c}\Delta$. For typical photolithographically defined mesa samples with dimensions of  $\sim 10 \mu\mathrm{m}$, $I_\mathrm{c}\sim 1mA$ - a dissipation of $\sim 1mW$ per junction. The effect of current injection on the superconducting gap has been considered previously using empirical models \cite{suzuki1}, resulting in the s-shaped re-entrant IV characteristics frequently observed.
\begin{figure}[h]
\begin{center}
\input epsf
\epsfxsize=75mm  \epsfbox{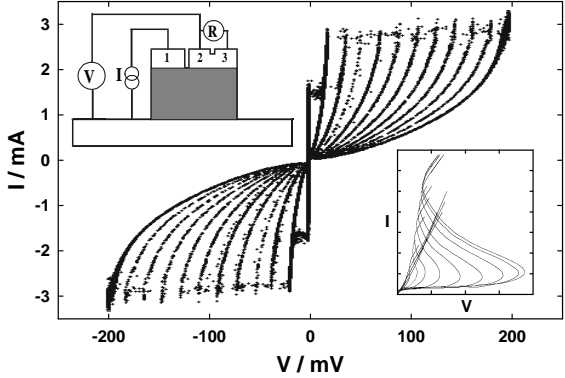} \vskip -10mm \caption{Hysteretic IV curve of an oxygen annealed sample of 11 junctions at a base temperature of 4.2K. Upper inset: Schematic geometry of mesa showing current contact (1), voltage contact (2). Sample temperature is monitored by measurement of resistance between 2 and 3. Lower inset: Re-entrant IV curves obtained over a range of temperature.} \label{schematic}
\end{center}
\end{figure}

In this paper we report measurements of the dc conductivity of small stacks of intrinsic junctions in 2212-BSCCO single crystals of varying oxygen stoichiometry. Two methods of minimising the effect of self-heating have been explored. A novel in-situ thermometer is used to measure the temperature of the mesa \emph{while taking dc measurements}, enabling corrections for heating to be made in the subsequent analysis. In order to reduce the self-heating, short pulses were used and simultaneous measurements of the IV characteristics and mesa temperature were made on a $\mu\mathrm{s}$ timescale. 
\section{Sample Fabrication}
Two as-grown crystals and two oxygen-annealed crystals were used in this study\cite{yang}. The crystals were fixed to single-crystal sapphire substrates using epoxy resin, and cleaved with adhesive tape until an optically smooth surface was obtained. Approximately 100nm of Ag was deposited onto the surface of each crystal by dc magnetron sputtering. The mesas were defined by contact photolithography and a combination of chemical etching and argon ion beam milling. Contacts were made to the mesas by means of tracks defined in a Au film $\sim1\mu \mathrm{m}$ thick. The sample geometry is shown in Figure 1 (upper inset). Four mesas were patterned on each crystal. Similar characteristics were observed for each mesa on a particular crystal.

To avoid interdiffusion of the contact and the crystal, the sample temperature was kept below 140\raisebox{1ex}{\scriptsize o}C throughout the fabrication process. The contacts display a well-defined and reproducible semiconductive temperature dependence, with a specific contact resistivity of $\sim 10^{-5}\Omega\mathrm{cm}^2$ at liquid $^4$He temperature. We were therefore able to use the contacts as a novel resistance thermometer to measure the mesa temperature during the acquisition of both dc and short-pulse IV characteristics. This allowed the characteristics to be corrected for heating.
\section{Results}
\subsection{DC Conductivity Measurements}
Each mesa displayed a highly hysteretic multiple-branched structure at low temperatures, as shown in figure 1, consistent with a series array of intrinsic junctions shunted by a significant capacitance and quasiparticle conductance. Each branch corresponds to a different number of junctions in the resistance state. The degree of hysteresis is determined by the value of $\beta _{c}$, which is calculated from a geometric estimate of the junction capacitance to be of order $10^3$. For junctions with sufficiently large capacitance, the resistive part of the characteristic is dominated by the quasiparticle conductance, except very close to the retrapping current. Measurements of the IV characteristics therefore allow the voltage and temperature dependence of the quasiparticle conductivity to be determined over a range of temperature\cite{lt22} below $T_\mathrm{c}$.
\begin{figure}[!h]
\begin{center}
\input epsf
\epsfxsize=75mm  \epsfbox{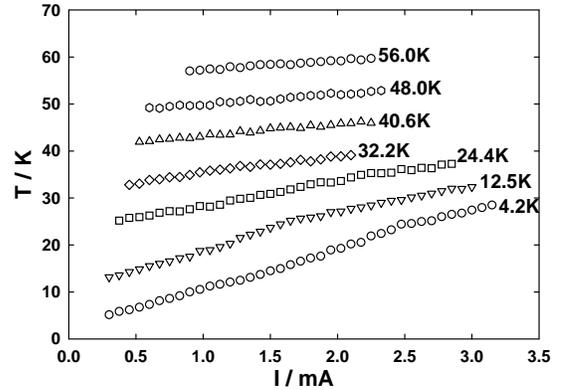} \vskip 0mm \caption{Mesa temperature plotted as a function of bias current for a typical mesa over a range of base temperatures.} \label{tempcurves}
\end{center}
\end{figure}

Measurements of the dc conductivity were performed by starting with the mesa in the zero-resistance state and slowly increasing the bias current until one of the junctions was driven into the resistive state. The current could then be increased further until the critical current of the next junction was exceeded, or decreased until the junction's retrapping current was reached in order to explore the voltage dependence of the conductivity. The IV characteristic was recorded using high-precision multimeters. The mesa temperature was measured simultaneously using the contact resistance between 2 and 3 (see figure 1 upper inset) with an ac current of $\sim1\mu \mathrm{A}$ and lock-in detection. Figure 2 shows the temperature of a typical mesa over a range of base temperature and bias current for the first resistive branch. Significant heating is observed at all temperatures. 

Characteristics were measured from a base temperature of 4.2K to close to $T_\mathrm{c}$ and were corrected for self-heating by interpolation. In figure 3, the solid and dashed lines are direct measurements of the first and 20th (normalised) resistive branches of a typical mesa at a base temperature of 28K. The dotted line shows the intrinsic IV characteristic at 28K deduced from measurements at lower base temperatures corrected for self-heating. Once corrected for heating, the characteristics for a single and 20 junctions lie on a universal curve, giving some confidence in the temperature correction technique. The uncorrected measured characteristics are re-entrant at high bias, whereas the corrected characteristics are significantly more linear. The divergent conductivity has previously been interpreted as the signature of the energy gap\cite{yurgens}(V=2$\Delta $) or as a nonequilibrium effect\cite{suzuki1}. Figure 3 shows that the effect is predominantly due to simple heating.
\begin{figure}[h]
\begin{center}
\input epsf
\epsfxsize=75mm  \epsfbox{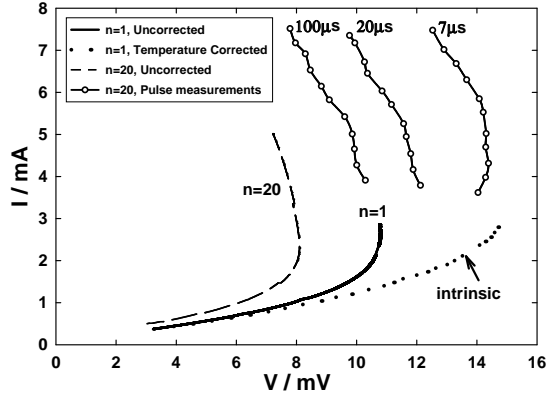} \vskip -10mm \caption{IV curves obtained at T=28K for a mesa of 20 junctions. Curves are shown for one junction in the resistive state (n=1), and all junctions resistive (n=20) scaled by a factor 1/20. Typical short pulse measurements are also shown for n=20.} \label{28kcurves}
\end{center}
\end{figure}
\subsection{Short Pulse Measurements}
The technique described above allows the intrinsic conductivity to be obtained from dc measurements (see section 3.3). However, much of the interest of these measurements is in the behaviour close to the energy gap, and high bias currents are required. The resulting dissipation, and consequent heating, is sufficiently large that dc measurements can not be used to explore this region, and short pulse measurements are necessary.
\begin{figure}[h]
\begin{center}
\input epsf
\epsfxsize=75mm  \epsfbox{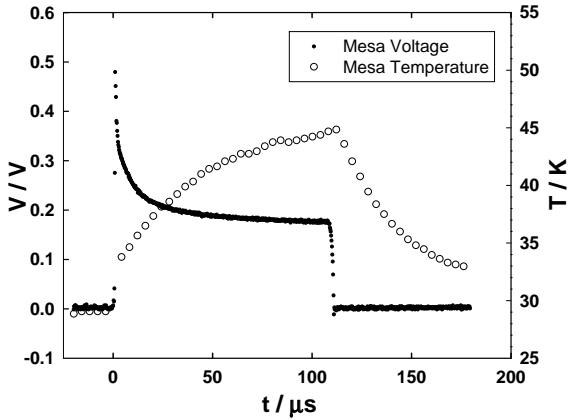} \vskip -10mm \caption{Voltage and temperature variation of mesa of 20 junctions when a current pulse of 6mA is applied at 28K.} \label{pulseVT}
\end{center}
\end{figure}
We have performed IV measurements on several samples using pulses of $\sim 100 \mu \mathrm{s}$ - 100ms, recording the mesa temperature by performing a simultaneous measurement of the temperature-dependent contact resistance by application of a small dc bias. The characteristics were obtained using a computer-controlled fast digital storage oscilloscope over a large range of temperature. These measurements reveal two distinct thermal time constants, $\tau_{mesa} \sim \mu \mathrm{s}$ corresponding to heating of the mesa relative to the crystal, and $\tau_{xtal} \sim \mathrm{ms}$ corresponding to heating of the sample above the base temperature. The time evolution of the IV characteristics is demonstrated in figure 4. At t=0 (the rising edge of the current pulse), the temperature of the mesa is the same as the thermal bath. At t=0 the mesa temperature begins to rise above that of the bath and the voltage across the junction begins to fall due to the reduction of $\rho_{c}$ with temperature. The points in figure 3 show the IV characteristics at different times after the rising edge of a given pulse. At large t, the curves approach the uncorrected dc characteristics. To obtain intrinsic characteristics by direct measurement the voltage must clearly be measured on a very short timescale, requiring a time resolution of better than $\sim 1 \mu \mathrm{s}$\cite{suzuki2}.
\subsection{Temperature Dependence of the Conductivity}
The conductivity of several samples was measured above $T_\mathrm{c}$ by application of a small ac bias. In addition, the quasiparticle conductivity was calculated in the superconducting state from the temperature-corrected IV characteristics described above. Some overlap in the temperature dependence was obtained by application of a large magnetic field (6.6T) in the c-direction to suppress Josephson coupling. The values obtained from the two methods in the overlap region are in good agreement. Figure 5 shows the conductivity of 3 samples from low temperatures to well above $T_\mathrm{c}$. Below $T_\mathrm{c}$, the ohmic component of the conductivity is plotted.

For all samples, $\sigma_{\mathrm{c}}(T)$ decreases
monotonically from temperatures well above $T_\mathrm{c}$, consistent with the onset of a pseudo-gap, possibly due to precursor pairing in the normal state. $\sigma_{c}$ is continuous at $T_\mathrm{c}$, with no evidence for the discontinuity in slope which might be expected from models involving the onset of pairing at $T_\mathrm{c}$. These data are in good quantitative agreement with the results of Latyshev and co-workers, who interpret their measurements in terms of a model involving a significant coherent contribution to the conductivity\cite{latyshev}.
\begin{figure}[h]
\begin{center}
\input epsf
\epsfxsize=75mm  \epsfbox{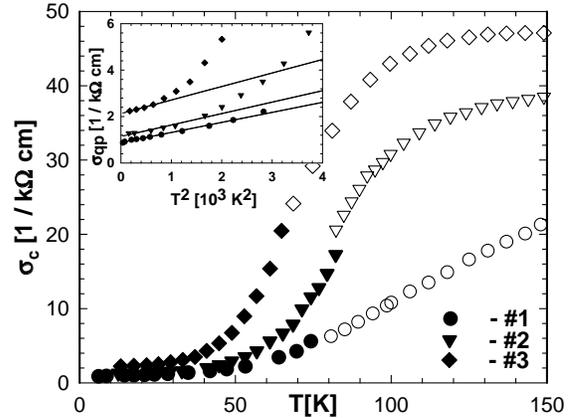} \vskip -10mm \caption{Quasiparticle conductivity plotted for three samples spanning
optimal doping (\#1 - $T_\mathrm{c}=75 \pm 3.5K$, \#2 -
$T_\mathrm{c}=87\pm0.9K$, \#3 - $T_\mathrm{c}=
86\pm0.5K$). Open
symbols - normal state. Solid symbols - from fits to IV curves
below $T_\mathrm{c}$. Inset shows low temperature behaviour.} \label{pulseVT}
\end{center}
\end{figure}
\section{Conclusions}
We have developed a novel method of measuring the temperature of BSCCO mesas which allows IV characteristics to be corrected for self-heating. This heating is shown to be significant over a large range of bias current, and is the main cause of the divergent conductivity which is often observed at high bias currents. Thermal gradients within the mesa itself may lead to an underestimate of the corrected mesa temperature. To explore the energy gap region with $V=2\Delta\sim$ 80mV per junction, pulses have to be used and measurements must be made on a sub-$\mu\mathrm{s}$ timescale.

Very similar results have been obtained for mesas on all four samples studied. Although some sample dependence might be expected, it is likely that self-heating will have been significant in all dc measurements on mesa structures previously published.
\section{Acknowledgments}
This work is supported by EPSRC. We thank Peter Andrews and Gary Walsh for technical support. PJT and JCF gratefully acknowledge financial support from EPSRC.

\end{document}